\documentclass[11pt]{article}
\usepackage{amsmath,amssymb}
\usepackage{epsfig}
\usepackage{mathrsfs}
\usepackage{amsmath,amssymb}
\usepackage{bbm}
\usepackage{bm}
\usepackage{epsfig}
\usepackage{graphicx}

\newcommand{\be}{\begin{equation}}
\newcommand{\ee}{\end{equation}}
\newcommand{\bea}{\begin{eqnarray}}
\newcommand{\eea}{\end{eqnarray}}
\newcommand{\unit}{\mathbbm{1}} 
\def\sqr#1#2{{\vcenter{\vbox{\hrule height.#2pt
        \hbox{\vrule width.#2pt height#1pt \kern#1pt
          \vrule width.#2pt}
        \hrule height.#2pt}}}}

\begin{document}

\begin{flushright}
SU-4252-857 \\
%SINP/TNP/XXXX \\
ITP-UH-15/07 \\
\end{flushright}

\begin{flushright}   
\today
\end{flushright} 
\begin{center}
\vskip 2em

{\Large \bf Interacting Quantum Topologies and the Quantum Hall Effect} \\

\vskip 1em

\centerline{ \bf A.~P.~Balachandran$^{a}$, Kumar~S.~Gupta$^{b}$, Se\c{c}kin~K\"{u}rk\c{c}\"{u}o\v{g}lu$^{c}$}

\vskip 1em

{\small
\centerline{\sl  $^a$ Department of Physics, Syracuse University, Syracuse NY 13244-1130 USA}

\vskip 1em

\centerline{\sl $^b$ Theory Division, Saha Institute of Nuclear Physics, 1/AF Bidhannagar, Kolkata 700064, India. }

\vskip 1em

\centerline{\sl  $^c$  Institut f\"ur Theoretische Physik, Universit\"at Hannover 
     Appelstra\ss{}e 2, D-30167 Hannover, Germany}

\vskip 1em

{\sl  e-mails:}  \hskip 2mm {\sl bal@phy.syr.edu, kumars.gupta@saha.ac.in, seckin.kurkcuoglu@itp.uni-hannover.de } }
\end{center}

\vskip 4em

\begin{abstract}
The algebra of observables of planar electrons subject to a constant background magnetic field $B$ is given by
${\cal A}_{\theta}(\mathbb{R}^{2}) \otimes {\cal A}_{\theta}(\mathbb{R}^{2})$ ($\theta = -\frac{4}{eB}$), the product of two mutually commuting Moyal algebras. It describes the free Hamiltonian and the guiding centre coordinates. We argue that ${\cal A}_{\theta}(\mathbb{R}^{2})$ itself furnishes a representation space for the actions of these two Moyal algebras, and suggest physical arguments for this choice of the representation space. We give the proper setup to couple the matter fields based on ${\cal A}_{\theta}(\mathbb{R}^{2})$ to electromagnetic fields which are described by the abelian commutative gauge group ${\cal G}_{c}(U(1))$, i.e. gauge fields based on ${\cal A}_0(\mathbb{R}^{2})$. This enables us to give a manifestly gauge covariant formulation of integer quantum Hall effect (IQHE). Thus, we can view IQHE as an elementary example of interacting quantum topologies, where matter and gauge fields based on algebras ${\cal A}_{\theta^\prime}$ with different $\theta^\prime$ appear. Two-particle wave functions in this approach are based on ${\cal A}_{\theta}(\mathbb{R}^{2}) \otimes {\cal A}_{\theta}(\mathbb{R}^{2})$. We find that the full symmetry group in IQHE, which is the semi-direct product $SO(2) \, \ltimes \, {\cal G}_{c}(U(1))$ acts on this tensor product using the twisted coproduct $\Delta_\theta$. Consequently, as we show, many particle sectors of each Landau level have twisted statistics. As an example, we find the twisted two particle Laughlin wave functions.
\end{abstract}

\vskip 3cm
\begin{quote}
\end{quote}

\newpage

\setcounter{footnote}{0}

\section{Introduction: Interacting Quantum Topologies}

Classical mechanics on a given spacetime $Q$ is formulated using the commutative algebra of smooth functions $C^{\infty}(T^*Q)$ on the phase space associated to $Q$. Classical topology is encoded in this algebra. In the passage to quantum theory, this algebra is deformed to an appropriate noncommutative algebra. If classical topology may be identified with the commutative algebra $C^{\infty}(T^*Q)$, then ``quantum topology'' can be identified with its noncommutative deformation.

Consider the flat spacetime ${\mathbb R}^{d+1}$. Its topology is encoded in the commutative algebra of smooth functions $C^\infty({\mathbb R}^{d+1}) \equiv {\cal {A}}_0({\mathbb R}^{d+1})$. A deformation of this algebra is the Moyal algebra ${\cal {A}}_{\theta}({\mathbb R}^{d+1})$. It is generated by the coordinate functions $x_0, x_1, \cdots, x_d$ satisfying 
\be
\lbrack x_\mu \,, x_\nu \rbrack_\star := x_\mu \star x_\nu - x_\nu \star x_\mu = i \theta_{\mu \nu} \,.
\ee 
The $\star$-product is defined by
\be
\alpha \star_\theta \beta = \alpha \, e^{\frac{i}{2} \theta_{\mu \nu} \overleftarrow{\partial}_\mu \overrightarrow{\partial}_\nu} \, \beta \,, \quad \alpha, \beta \in {\cal {A}}_{\theta}(R^{d+1}) \,.
\ee
Much work has recently been done on the formulation of quantum field theories using this algebra which may then be identified as the noncommutative or quantum topology of spacetime \cite{Balachandran:2005eb, Balachandran:2006pi, Balachandran:2006ib}.

Now, a theory where different sorts of matter and gauge fields are based on algebras ${\cal A}_\theta({\mathbb R}^{d+1})$ with different $\theta$ is an example of a theory of interacting quantum topologies. Theories with such interacting topologies have been developed in \cite{Balachandran:2006ib}. In this paper we shall argue that the integer quantum Hall effect (IQHE) provides a concrete example of such a system.

As it is well known, IQHE studies the properties of non-relativistic planar electrons subject to a constant background magnetic field $B$. The algebra of observables in IQHE is given by the product of two mutually commuting Moyal algebras ${\cal A}_{\theta}(\mathbb{R}^{2}) \otimes {\cal A}_{\theta}(\mathbb{R}^{2})$ ($\theta = -\frac{4}{eB}$). These Moyal algebras describe the single particle free Hamiltonian and the guiding centre coordinates.

To form the Lagrangian or Hamiltonian it is necessary to multiply the fields. Quantum free fields are superpositions of single particle wave functions with operator coefficients. Thus in the passage to quantum field theory and multiparticle dynamics, we assume that wave functions can be multiplied with each other and hence form elements of an algebra. (More precisely it is the space of test functions of quantum fields which form an algebra). In conventional treatment of IQHE, this algebra is the commutative algebra ${\cal A}_0(\mathbb{R}^{2})$. It is the algebra appropriate for $\mathbb{R}^{2}$ as the spatial slice. Thus the algebra associated with the wave function reflect the spatial topology.

However, with only finitely many Landau levels accessible, the algebra of observables is Moyal algebra times a finite-dimensional matrix algebra. The latter are formed of matrices acting on the finite-dimensional vector space of accessible Landau levels. From this algebra of observables it is not possible to recover the algebra of smooth functions ${\cal A}_0(\mathbb{R}^{2})$. Thus we have only the guiding centre algebra to probe the spatial location in IQHE. We must accept it as the effective spatial algebra for IQHE. 

It follows that the algebra of wave functions, which describe the properties of the spatial slice, is appropriately chosen as the Moyal algebra ${\cal A}_{\theta}(\mathbb{R}^{2})$.

Guided by these considerations, in this paper we assume that the spatial slices are described by the algebra ${\cal A}_\theta(\mathbb{R}^{2})$ of the guiding centre and demonstrate (as is necessary for consistency) that ${\cal A}_{\theta}(\mathbb{R}^{2})$ itself furnishes a representation space for the actions of the observables of IQHE. We give the proper setup to couple the matter fields based on ${\cal A}_{\theta}(\mathbb{R}^{2})$ to electromagnetic fields which are of course described by the abelian commutative gauge group ${\cal G}_{c}(U(1))$. This enables us to give a manifestly gauge covariant formulation of IQHE. Thus, we can view IQHE as an elementary example of interacting quantum topologies, where different sorts of matter and gauge fields based on algebras ${\cal A}_{\theta^\prime}$ with different values of $\theta^\prime$ appear.

Single particle wave functions in this approach are the same as those that come out of the standard treatment, but at the level of many particle states and/or quantum field theory, there is indeed a difference. $N$-particle wave functions are based on the $N$-fold tensor product $\otimes_N {\cal A}_{\theta}(\mathbb{R}^{2})$. We find that the full symmetry group in IQHE, which is the semi-direct product $SO(2) \, \ltimes \, {\cal G}_{c}(U(1))$, acts on such tensor products by the twisted coproduct\footnote{The concept of Drinfel'd twisted symmetries was first introduced in the context of Groenewold-Moyal spacetimes in \cite{Chaichian:2004za, Wess:2003da, Aschieri:2005yw}, where it was utilized to define twisted Poincar{\'e} and diffeomorphism groups as symmetries on ${\cal A}_{\theta}(\mathbb{R}^{d+1})$.} $\Delta_\theta$. Consequently, we find that the $N$-particle sectors of each Landau level have twisted statistics [See, \cite{Balachandran:2005eb, Balachandran:2006pi} for a detailed discussion of twisted statistics on ${\cal A}_{\theta}(\mathbb{R}^{d+1})$]. As an example, we explicitly work out the the twisted two particle Laughlin wave functions.

\section{Planar Electrons}

Consider a single electron in a plane, moving in a magnetic field $B$ perpendicular to the plane. In the symmetric gauge, the gauge potential for the magnetic field is given by
\be
A_a = \frac{B}{2} \epsilon_{ab} y_b,
\ee
where $y_a$ denote the coordinate on the plane with $a,b = 1,2$ and sum over the repeated indices is assumed. The Lagrangian for the electron of mass $m$ moving in the plane in the presence of the field $B$ is given by
\be
L = \frac{1}{2} m {\dot{y_a}}^2 - \frac{eB}{2} \epsilon_{ab}{\dot{y_a}}y_b
\ee
The momentum conjugate to the coordinate $y_a$ is given by
\be
p_a = m {\dot{y_a}} - \frac{eB}{2} \epsilon_{ab}y_b.
\ee
We can equivalently write
\bea
{\dot{y_a}} &=& \frac{eB}{2m} \epsilon_{ab}\xi_b \,, \\
\xi_b &\equiv& y_b + \frac{2}{eB} \epsilon_{cb}p_c \,.
\eea
$\xi_a's$ fulfill
\be
[\xi_a , \xi_b] = - i \theta \epsilon_{ab}, ~~~ \theta = - \frac{4}{eB}.
\ee
The Hamiltonian is given by 
\be
H = \frac{1}{2} m {\dot{y_a}}^2 = \frac{1}{2m} \left ( \frac{eB}{2} \right )^2 \xi_a^2 \,.
\label{eq:Hamiltonian}
\ee
It describes a harmonic oscillator with the energy eigenvalues $E = \left ( n + \frac{1}{2} \right ) \omega$ with 
$\omega = \frac{e B}{m} $. 

The guiding centre coordinates for the electron are defined by 
\be
X_a = y_a - \frac{2}{eB} \epsilon_{ba} p_b \,,
\ee
and they satisfy the commutation relations
\be
[X_a, X_b] = i \theta \epsilon_{ab} \,.
\ee
It is easy to see that
\be
\lbrack X_a, \xi_b \rbrack = 0 \,.
\ee
Hence, $X_a$ and $\xi_a$ form a basis for two mutually commuting Moyal algebras which we will denote by ${\cal A}_\theta^L$ and ${\cal A}_\theta^R$ respectively from now on. 

An irreducible representation space for the actions of $\xi_a$ and $X_a$ is ${\cal A}_{\theta}(\mathbb{R}^{2})$. It has the $\star$-product:
\be
f \star g = f e^{\frac{i}{2} \theta \varepsilon_{\mu \nu} \overleftarrow{\partial}_\mu \overrightarrow{\partial}_\nu} g \,, \quad 
f \,, g \in {\cal A}_{\theta}(\mathbb{R}^{2}) \,.
\ee
$X$ and $\xi$ act on ${\cal A}_\theta(\mathbb{R}^{2})$ by left and right $\star$ multiplication, respectively. Thus for any $\alpha \in {\cal A}_\theta(\mathbb{R}^{2})$ we have
\bea
X_a \alpha &=& x_\alpha \star \alpha \,, \\
\xi_a \alpha &=& \alpha \star x_\alpha \,.
\eea
Thus, ${\cal A}_\theta(\mathbb{R}^{2})$ is a left module under the action $\rho({\cal A}_\theta^L)$ and right module under the action of $\rho({\cal A}_\theta^R)$, where $\rho$ denotes representations of $({\cal A}_\theta^{L,R})$. 

We emphasize that we treat ${\cal A}_\theta(\mathbb{R}^{2})$ as a module algebra : besides carrying the the representation of the observables, it is itself an algebra. we will have further comments on this approach and its comparison with the standard approach below.

For the  formulation of quantum theory, we need also a scalar product on ${\cal A}_\theta(\mathbb{R}^{2})$, the completion of ${\cal A}_\theta(\mathbb{R}^{2})$ in this scalar product giving the Hilbert space ${\cal H}$ of the theory. This scalar product can be taken to be the conventional one:
\be
(\alpha, \beta) = \int d^2 x {\bar \alpha}(x) \beta(x) \,, \quad \alpha \,, \beta \in {\cal A}_\theta(\mathbb{R}^{2}) \,.
\ee
$X_a$ and $\xi_b$ are self-adjoint for this scalar product, as they should be.

{\it Remark:} ${\cal A}_{\theta}(\mathbb{R}^{2})$ is the algebra of operators on $L^2({\mathbb R}^2)$, the algebra generated by a position $x$ and a momentum $p$. It is spanned by $|m \rangle \langle n |$ where $|m\rangle$ for example is the $m^{th}$ harmonic oscillator 
state. If $N$ Landau levels are filled, we can restrict $n$ to $n \leq N$.

Before closing this section, we note that there is no difference between our treatment and the standard one based on ${\cal A}_{0}(\mathbb{R}^{2})$ for single particle dynamics. This will be explicitly verified in section $4.1$.

\section{Quantum Hall System as a Model of Interacting Quantum Topologies}

In QHE on a plane, besides translational symmetry electrons possess rotational symmetry about the perpendicular axis. $SO(2)$ spatial rotations 
acts on $X_{a}$, $\xi_a$ in the usual way:
\be
X_a \rightarrow {\cal R}_{ab} X_b \,, \quad \xi_a \rightarrow {\cal R}_{ab} \xi_b \,, \quad {\cal R} \in SO(2) \,. 
\ee
This action is an automorphism of ${\cal A}_{\theta}(\mathbb{R}^{2})$ since,
\be
{\cal R} \varepsilon {\cal R}^{T}= \varepsilon \,,
\ee
and we can explicitly check for example that:
\be
[{\cal R}_{ac} X_{c}, {\cal R}_{bd} X_{d}] = i \theta {\cal R}_{ac} {\cal R}_{bd} \epsilon_{cd} = i \theta \epsilon_{ab} \,.
\ee

What is the coproduct $\Delta$ on SO(2)? If $\mu_{\theta}$ is the multiplication map on ${\cal A}_{\theta}(\mathbb{R}^{2})$,
then
\be
\mu_{\theta} (f \otimes g) = \mu_{0} \left ( {\cal F}_{\theta}^{-1} f \otimes g \right ) = f \star g,
\ee
where
\be
{\cal F}_{\theta}^{-1} = e^{\frac{i}{2} \partial_{\mu} \otimes \theta^{\mu \nu} \partial_{\nu}},
\ee
and
\be
m_{0}(f \otimes g) = f \cdot g \,, 
\ee
is the pointwise multiplication in ${\cal A}_{0}(\mathbb{R}^{2})$.

In order for $\Delta$ to be consistent with the multiplication map on ${\cal A}_{\theta}(\mathbb{R}^{2})$, it must satisfy
\be
\mu_{\theta} \left ( \Delta({\cal R}) (f \otimes g) \right ) = {\cal R} (f \star g) \,.
\label{eq:twistedR}
\ee

Let us now note that the $SO(2)$ invariance of $\varepsilon$ implies that any coproduct $\Delta_{\theta^{'}}$ satisfying
\be
\Delta_{\theta^{'}}({\cal R}) = {\cal F}_{\theta^{'}}({\cal R} \otimes {\cal R}) {\cal F}_{\theta^{'}}^{-1},
\ee
fulfills (\ref{eq:twistedR}).
 
So the natural question that arises is: how can we fix $\theta^{'}$?. This question can be 
successfully answered by studying the gauge symmetry of the system.

In QHE, in addition to the $B$-field perpendicular to the plane, we need to couple the system to an electric field in the plane (at least to see the Hall current).These fields couple to the commuting coordinate $y_a$, with the conventional covariant derivative,
\be
{\tilde D}_a = \frac{\partial}{\partial y_a} + ie \, A_a(y) + ie \, S_a(y) \,.
\ee
Here $A_a$ is the gauge potential generating the perpendicular magnetic field $B$,
\be
A_a = \frac{B}{2} \epsilon_{ab} \, y_b \,,
\ee
as defined previously, and $S_a$ is the additional potential in the $1-2$ plane. 

${\tilde D}_a $ transform under the commutative gauge group ${\cal G}_{c}(U(1))$ as
\be
\tilde{D}_a \rightarrow {\cal U}(y) \tilde{D}_a {\cal U}(y)^{-1} \,. 
\ee
It follows that we can keep $A_a$ fixed and transform $D_a:={\tilde D}_a - i e A_a(y)$ as follows:
\be
D_a := \frac{\partial}{\partial y_a} + ie S_a \rightarrow {\cal U}(y) D_a {\cal U}(y)^{-1} \,.
\ee

We can now write down the action of $D_a$ on ${\cal A}_{\theta}(\mathbb{R}^{2})$. We can express the commutative coordinate as
\cite{Balachandran:2006ib}
\be
y_a= \frac{1}{2} (X_a+\xi_a) \,.
\ee
Therefore, if $\phi \in {\cal A}_{\theta}(\mathbb{R}^{2})$ is a charged field, it transforms under ${\cal G}_{c}(U(1))$ as
\be
\phi(x) \rightarrow {\cal U} (X_a+\xi_a) \phi (x) \,.
\ee
Here
\be
(X_a+\xi_a) \phi(x) = \hat{x}_a \star \phi(x) + \phi(x) \star \hat{x}_a = 2 x_a \varphi(x) \,,
\ee
where
\be
\hat{x}_a(x)=x_a \,.
\ee

Thus the covariant derivative of $\phi(x)$ is given as
\be
\partial_\mu \phi (x) + ie (A_{\mu} \phi)(x)= \partial_\mu \phi(x)+ ie A_\mu(x) \phi (x) \,.
\ee

It is necessary that, we couple quantum fields $\psi_{\theta}$ with $U(1)$ charge to construct interactions. We should be able to form a charge-neutral Hamiltonian and charge densities by multiplying $\psi_{\theta}$'s and $\bar{\psi}_{\theta}$'s.
Thus, if $\psi_\theta$ is a charged field, we have to know how to consistently gauge transform $\bar{\psi_{\theta}} \star \psi_{\theta}$, $\psi_{\theta} \star \psi_{\theta}$ etc.

The requirement that
\be
\mu_{\theta} \, (\Delta ({\cal U}) \psi_{\theta} \otimes \psi_{\theta}) = {\cal  U} (\psi_{\theta} \star \psi_{\theta})
\ee
uniquely fixes the coproduct $\Delta$ for the gauge group as
\be
\Delta = \Delta_{\theta} \,, \quad \Delta_{\theta}({\cal U}) = {\cal F}_{\theta}({\cal U} \otimes {\cal U}) {\cal F}_{\theta}^{-1} \,.
\ee

We can now go back to the implementation of $SO(2)$ rotations on ${\cal A}_{\theta}(\mathbb{R}^{2})$. The group $SO(2)$ acts on ${\cal G}_{c}(U(1))$. Therefore, the full group is the semi-direct product
\be
SO(2) \, \ltimes \, {\cal G}_{c}(U(1)) \,. 
\ee
Thus, the coproduct must preserve this group structure. Consequently, if $\Delta_{\theta}$ is the coproduct for 
${\cal G}_{c}(U(1))$, it is the same for $SO(2)$:
\be
\Delta_{\theta}({\cal R}) = {\cal  F}_{\theta} ({\cal R} \otimes {\cal R}) {\cal F}_{\theta}^{-1} \,.
\ee

Thus, the preceding discussion gives a manifestly gauge covariant formulation of integer QHE with a unique choice of the coproduct.

\section{Twisted Laughlin States}

\subsection{Landau Levels}

The eigenstates for the single particle Hamiltonian (\ref{eq:Hamiltonian}) are the Landau levels and they can easily be constructed in the preceding formalism. We give a very brief account of this for completeness. 

We start with forming the complex combinations
\bea
&&\xi = \xi_1 + i \xi_2 \,, \quad \bar{\xi} = \xi_1 - i \xi_2 \,, \\
&&X = X_1 + i X_2 \,, \quad \bar{X} = X_1 - i X_2 \,.
\eea
The lowest Landau level (LLL) condition for a wave function $\alpha \in {\cal A}_{\theta}(\mathbb{R}^{2})$ is then given as
\be
{\bar \xi} \, \alpha = \alpha \star {\bar z} = 0 \,,
\ee
where ${\bar z} = x_1 - i x_2$. Expanding the star product we find the first order differential equation
\be
\left ( \partial_z + \frac{2}{\theta} {\bar z} \right ) \alpha (z, {\bar z}) = 0 \,, 
\ee
whose solutions are of the form 
\be
\alpha_0(z, {\bar z}) = \lambda({\bar z}) e^{-\frac{2}{\theta} |z|^2}\,,
\ee
where $\lambda$ is an arbitrary analytic function. 

Higher Landau levels can be constructed by acting on $e^{-\frac{2}{\theta} |z|^2}$ by powers of $\xi$ and ${\bar X}$. In this way the $n^{th}$ Landau level is given by
\be
{\bar X}^n \xi^n \, e^{-\frac{2}{\theta} |z|^2} = {\bar z}^n \star e^{-\frac{2}{\theta} |z|^2} \star z^n 
= C_n L_n(z, {\bar z}) e^{-\frac{2}{\theta} |z|^2} \,,
\ee 
where $L_n$ is the Laguerre polynomial and $C_n$ is a constant.

The levels with fixed angular momentum in each Landau level can also be determined. In LLL the $0$ angular momentum state is given by the condition
\be
X \alpha_0 (z, {\bar z}) = z \star \alpha_0 (z, {\bar z}) = 0 \,,
\ee
which is solved by $\lambda({\bar \xi}) = constant$. The state with angular momentum $\ell$ is then constructed by 
\be
{\bar X}^\ell \alpha_0(z, {\bar z}) = \underbrace{{\bar z} \star {\bar z} \cdots \star {\bar z}}_{\ell-times} \star \alpha_0
(z, {\bar z}) = {\bar z}^\ell \star \alpha_0 (z, {\bar z})  = D_\ell  {\bar z}^\ell \alpha_0 (z, {\bar z}) \,, 
\ee
where $D_\ell$ is a constant.

States with fixed angular momentum in higher Landau levels are constructed in a similar manner. We observe that all these results are in agreement with those of the standard treatments \cite{Ezawa}.

\subsection{Twisted Statistics} 

The total symmetry group $SO(2) \, \ltimes \, {\cal G}_{c}(U(1))$ acts on the products of wave functions by the twisted coproduct
$\Delta_\theta$. Consequently, many particle states must have twisted statistics. Let us see how this comes about \cite{Balachandran:2005eb} \cite{Balachandran:2006pi}. (See also the earlier work of Oeckl \cite{Oeckl:2000eg}.)

The space of single particle wave functions are elements of the module algebra ${\cal A}_\theta$. Hence the space of $n$-particle
wave functions is associated with the $n$-fold tensor product ${\cal A}_\theta \otimes {\cal A}_\theta \otimes\cdots
\otimes {\cal A}_\theta$. 

Let us be more concrete. Let $\chi_1, \chi_2 \in {\cal A}_\theta$ and consider $\chi_1 \otimes \chi_2$ and $\chi_2 \otimes \chi_1$. The standard flip map is defined by $\tau_0(\chi_1 \otimes \chi_2) = \chi_2 \otimes \chi_1$. But, $\tau_0$ does not commute with $\Delta_\theta$ and thus it is not possible to use it to construct irreducible subspaces
of $\Delta_\theta$. However, it can be shown that the twisted flip operator $\tau_\theta = {\cal F}_{\theta} \tau_0 {\cal F}_{\theta}^{-1} 
= ({\cal F}_{\theta}^{-1})^2 \tau_0 $ satisfies
\begin{equation}
\lbrack \tau_\theta \,, {\cal F}_{\theta} \Delta(g) {\cal F}_{\theta}^{-1} \rbrack  =  0 \,, \quad \tau_\theta^2 \ =\  \unit \,.
\label{eq:twistedflip}
\end{equation}
where $g$ is an element of $SO(2) \, \ltimes \, {\cal G}_{c}(U(1))$. Assuming that $\tau_\theta$ is superselected (as is the case for $\theta =0$), we infer from (\ref{eq:twistedflip}) that the irreducible subspaces for $\Delta_\theta(g)$ are given by
\begin{equation}
{\cal \chi}_\theta^\pm = \frac{1 \pm \tau_\theta}{2} ({\cal \chi}_1 \otimes {\cal \chi}_2) \,.
\label{eq:irrsubspaces}
\end{equation}
These subspaces define the generalized bosons and fermions for the upper and the lower signs, respectively.

\subsection{Twisted Laughlin Wave Functions}

In the ordinary formulation of IQHE with filling factor $\nu = 1$, the Laughlin wave functions for $N$-particle state is given by the totally antisymmetric tensor product of the single particle wave functions at positions $z_1, \cdots, z_N$ and with angular momentum taking the values $\ell = 0 \cdots N-1$. 

Let us now see how $N=2$ state is twisted by applying the results of the previous subsection. Let us use the convention that 
the first slot in the two fold tensor product of single particle states belongs to the first particle and the second slot belongs to the second particle. Next, following \cite{Balachandran:2006ib}, we introduce these labels into the Drinfel'd twist element: 
\begin{equation}
{\cal F}_{\theta}^{-1} = e^{\frac{1}{2} \theta (\partial_{z_1} \otimes \partial_{{\bar z}_2} - \partial_{{\bar z}_1} \otimes \partial_{z_2})} \,.
\end{equation}

Now we compute the twisted two particle state. Given that the single particle states
(up to overall normalizations) are the angular momentum $0,1$ Laughlin states $e^{-\frac{2}{\theta}|z|^2}$ and ${\bar z} \, e^{-\frac{2}{\theta}|z|^2}$ respectively, we have, for the twisted two particle state,

\begin{equation}
\frac{1}{\sqrt{2}} \left (e^{-\frac{2}{\theta} |z_1|^2} \otimes {\bar z}_2 e^{-\frac{2}{\theta} |z_2|^2} - {\cal F}^{-2}_\theta {\bar z}_1 e^{-\frac{2}{\theta} |z_1|^2} \otimes e^{-\frac{2}{\theta} |z_2|^2} \right )
\label{eq:twisted2p}
\end{equation}
The second term in this expression can be computed rather easily. We have 
\bea
{\cal F}_\theta^{-2} {\bar z}_1 e^{-\frac{2}{\theta} |z_1|^2} \otimes e^{-\frac{2}{\theta} |z_2|^2} &=& -\frac{\theta}{2} {\cal F}_\theta^{-2} ( \partial_{z_1} \otimes 1) e^{-\frac{2}{\theta} |z_1|^2} \otimes e^{-\frac{2}{\theta} |z_2|^2} \nonumber \\
&=& -\frac{\theta}{2}(\partial_{z_1} \otimes 1) {\cal F}_\theta^{-2} e^{-\frac{2}{\theta} |z_1|^2} \otimes e^{-\frac{2}{\theta} |z_2|^2} \nonumber \\
&=& - \frac{\theta}{2} (\partial_{z_1} \otimes 1) e^{\frac{4}{\theta} ({\bar z}_1 \otimes z_2 - z_1 \otimes {\bar z}_2)} e^{-\frac{2}{\theta} |z_1|^2} \otimes e^{-\frac{2}{\theta} |z_2|^2}  \\
&=& ({\bar z}_1 \otimes 1 + 1 \otimes 2 {\bar z}_2) e^{ \frac{4}{\theta} ({\bar z}_1 \otimes z_2 - z_1 \otimes {\bar z}_2)} e^{-\frac{2}{\theta} |z_1|^2} \otimes e^{-\frac{2}{\theta} |z_2|^2} \nonumber \,.
\eea
where in passing from the second line to the third line we have used the fact that $\partial_{z_1} \otimes 1$  and ${\cal F}_\theta^{-2}$ commutes. Inserting this back into (\ref{eq:twisted2p}) we find  
\begin{equation}
\left ((1 \otimes {\bar z}_2) - ({\bar z}_1 \otimes 1 + 1 \otimes 2 {\bar z}_2) e^{ \frac{4}{\theta} ({\bar z}_1 \otimes z_2 - z_1 \otimes {\bar z}_2)} \right ) e^{-\frac{2}{\theta} |z_1|^2} \otimes e^{-\frac{2}{\theta} |z_2|^2} \,.
\label{eq:un1}
\end{equation}

Consider now the wave function for two angular momentum zero states. We have
\begin{equation}
\frac{1}{\sqrt{2}} \left (e^{-\frac{2}{\theta} |z_1|^2} \otimes e^{-\frac{2}{\theta} |z_2|^2} - {\cal F}^{-2}_\theta e^{-\frac{2}{\theta} |z_1|^2} \otimes e^{-\frac{2}{\theta} |z_2|^2} \right ) \,.
\label{eq:twisted2psame}
\end{equation}
Without the ${\cal F}^{-2}_\theta$ this two particle state would have vanished; whereas now we have,
\be
{\cal F}_\theta^{-2} e^{-\frac{2}{\theta} |z_1|^2} \otimes e^{-\frac{2}{\theta} |z_2|^2} = e^{\frac{4}{\theta} ({\bar z}_1 \otimes z_2 - z_1 \otimes {\bar z}_2)} e^{-\frac{2}{\theta} |z_1|^2} \otimes e^{-\frac{2}{\theta} |z_2|^2} \,.
\ee
So for (\ref{eq:twisted2psame}) we find
\begin{equation}
\left (1- e^{ \frac{4}{\theta} ({\bar z}_1 \otimes z_2 - z_1 \otimes {\bar z}_2)} \right ) e^{-\frac{2}{\theta} |z_1|^2} \otimes e^{-\frac{2}{\theta} |z_2|^2} \,.
\label{eq:un2}
\end{equation}

The wave functions in (\ref{eq:un1}) and (\ref{eq:un2}) are not normalized. But, normalization constants can be computed in a straightforward manner.

Finally, we note that although states like ${\cal F}_{\theta}^{-1} \left ( \frac{1 - \tau_0}{2} \right ) (\chi_1 \otimes \chi_2)$ are eigenstates of $\tau_\theta$ for eigenvalue $-1$, the map $\chi_1 \otimes \chi_2 \rightarrow
{\cal F}_{\theta}^{-1} \left ( \frac{1 - \tau_0}{2} \right ) (\chi_1 \otimes \chi_2)$ does not give states transforming under $\Delta_\theta({\cal R})$. That is because ${\cal F}_{\theta}^{-1} \left ( \frac{1 - \tau_0}{2} \right )$ does not commute with $\Delta_\theta({\cal R})$:
\be 
{\cal F}_{\theta}^{-1} \left ( \frac{1 - \tau_0}{2} \right ) \, \Delta_\theta ({\cal R}) \neq
\Delta_\theta ({\cal R}) \, {\cal F}_{\theta}^{-1} \left ( \frac{1 - \tau_0}{2} \right ) \,.
\ee

\section{Conclusion and Outlook}

In this paper we have shown that the Moyal plane ${\cal A}_{\theta}(\mathbb{R}^{2})$ itself furnishes an irreducible representation space for the actions of the algebra of observables in the IQHE. Using this result, we have given a manifestly gauge covariant formulation of IQHE. From the perspective of quantum field theory our results can be understood as follows. At the level of single particle wave functions our results are in complete agreement with the conventional treatment based on the commutative algebra ${\cal A}_{0}(\mathbb{R}^{2})$. However, at the level of multiparticles, or quantum field theory, our results indeed differ from that of the conventional treatments. This fact is manifestly reflected in the form of the multi-particle wave functions we have found in section 4.3. Possible effects of this new formulation on observable quantities such as energy and the Hall current, as well as the implementation of twisted discrete symmetries such as parity and time reversal in this system, are under investigation.

\section*{Acknowledgements}
Authors would like to thank P.Pre\v{s}najder for his kind and generous hospitality at Comenius University, Bratislava where a part of this work was done. A.P.B is supported by the DOE grant number DE-FG02-85ER40231. S.K. is supported by
the Deutsche Forschungsgemeinschaft (DFG) grant number LE 838/9.

\end{document}